\documentclass[aps,prb,showpacs,preprint,amssymb]{revtex4}
\usepackage{epsfig}

\begin{document}


\title{{\small JETP Letters, Vol. 75, No. 2, 2002, pp. 93-97.}\\[0.5cm]
Coherent Tunneling between Elementary Conducting Layers
in the NbSe$_3$ Charge-Density-Wave Conductor}

\author{Yu.~I. Latyshev}
\affiliation{Institute of Radio Engineering and Electronics,
Russian Academy of Sciences, Moscow 101999, Russia}
\author{A. A. Sinchenko}
\affiliation{Moscow State  Engineering-Physics Institute, Moscow 115409,
Russia}
\author{L. N. Bulaevskii}
\affiliation{Los Alamos National Laboratory, Los Alamos, New Mexico NM 87545, USA}
\author{V. N. Pavlenko}
\affiliation{Institute~of Radio Engineering and Electronics,
Russian Academy of Sciences, Moscow 101999, Russia}
\author{P.~Monceau}
\affiliation{Centre de Recherches sur les tres Basses Temperatures, BP 166,
38042 Grenoble, France}


\begin{abstract}

Characteristic features of transverse transport along the $a^*$
axis in the NbSe$_3$ charge-density-wave conductor are studied. At
low temperatures, the $I$-$V$ characteristics of both layered
structures and NbSe$_3$-NbSe$_3$ point contacts exhibit a strong
peak of dynamic conductivity at zero bias voltage. In addition, the
$I$-$V$ characteristics of layered structures exhibit a series of
peaks that occur at voltages equal to multiples of the double
Peierls gap. The conductivity behavior observed in the experiment
resembles that reported for the interlayer tunneling in Bi-2212
high-$T_c$ superconductors. The conductivity peak at zero bias is
explained using the model of almost coherent interlayer tunneling
of the charge carriers that are not condensed in the charge
density wave.

\end{abstract}

\maketitle

It is well known that the crystal structure of BSCCO high-$T_c$
superconductors consists of atomically thin superconducting
cuprate layers spatially separated by atomically thin insulating
layers of BiO and SrO. The corresponding spatial modulation of the
superconducting order parameter in the direction across the layers
(along the $c$ axis) leads to a discrete description of the
transverse transport (the Lawrence-Doniach model \cite{1}) with
the neighboring superconducting layers being coupled through
Josephson tunneling junctions. The validity of this approach is
confirmed by the results of the experiments on natural Bi-2212
layered structures of small lateral size \cite{2,3}. Currently,
the study of the interlayer tunneling of Cooper pairs and
quasiparticles \cite{4,5} is one of the new original methods for
investigating high-$T_c$ superconductors.

In this paper, we study the interlayer tunneling in a layered
system with an electron condensate of a different type, namely, in
a charge-density-wave (CDW) conductor. The material chosen for our
experiments is NbSe$_3$. This compound is characterized by two
Peierls transitions, which occur at the temperatures $T_{p1}=$145
K and $T_{p2}=$59 K. In the low-temperature Peierls state, the
Fermi surface retains some regions where the nesting conditions
are not satisfied (the "pockets") and, hence, the Peierls gap is
absent. Therefore, NbSe$_3$ does not undergo transition to the
insulating state and retains metallic properties down to the
lowest temperatures \cite{6}. From the analysis of both the
crystal structure of NbSe$_3$ and the anisotropy of its
properties, it follows that this material can be classed with
quasi-two-dimensional layered compounds. In fact, its conductivity
anisotropy in the ($b-c$) plane is determined by the chain
conductivity along the $b$ axis and is estimated as
$\sigma_b/\sigma_c\sim 10$, whereas the conductivity ratio
$\sigma_b/\sigma_{a^*}$ is determined by the layered character of
the structure and reaches the values $\sim10^4$ at low
temperatures \cite{7,8}. Figure 1a shows the crystal structure of
NbSe$_3$ in the ($a-c$) plane. One can see that, in this material,
the layered structure is formed as a result of a pairwise
coordination of selenium prisms with the predominant orientation
of their bases in the ($b-c$) plane. In Fig. 1b, the shaded areas
indicate the elementary conducting layers in which the prisms are
rotated and shifted with their edges toward each other. In these
layers, the distances between the niobium chains are relatively
small, whereas the neighboring conducting layers are separated by
an insulating layer formed as a double barrier by the bases of the
selenium prisms.

This type of layered structure in combination with the
conductivity anisotropy offers the possibility for the CDW
condensation in the elementary conducting layers spatially
separated by atomically thin insulating layers. In this case, as
in layered high-$T_c$ superconductors, one can expect that the CDW
order parameter will be modulated along the $a^*$ axis, and the
transport across the layers will be determined by the intrinsic
interlayer tunneling between the elementary layers with the CDW.

To verify these speculations, we carried out an experimental study
of the transport across the layers in NbSe$_3$ in the condensed
CDW state.

The samples used in the experiment were layered structures with a
small area through which the current flows across the layers, $S=
2\times 2$ $\mu$m$^2$, and with the number of elementary layers $\sim$30 (the
overlap junctions). They were made from a thin NbSe$_3$ single
crystal by the focused ion beam processing technique developed for
fabricating similar structures from Bi-2212 \cite{9}.
Schematically, the structure under investigation is shown in the
inset of Fig. 2. In addition, we studied the characteristics of
NbSe$_3$-NbSe$_3$ point contacts oriented in the direction of the
$a^*$ axis. The configuration of such a contact is shown in the
inset of Fig. 3. The contact was formed directly at low
temperature by bringing two NbSe$_3$ whiskers together with high
accuracy. In both cases, we performed four-terminal measurements
of the $I$-$V$ characteristics and their derivatives at low
temperatures, i.e., below the second Peierls transition
$T_{p2}=59$ K.

The main results of the experiment are shown in Fig. 2, which, for
one of the virtually perfect structures (No. 3), displays the
dynamic conductivity along the $a^*$ axis versus the bias voltage $V$
at different temperatures from 59.5 to 4.2 K. It can be seen that,
for $T<T_{p2}$, the differential $I$-$V$ characteristics are of a
tunneling character. When the temperature decreases below $\sim$35
K, the $I$-$V$ characteristics begin to exhibit a conductivity peak at
zero bias voltage. As the temperature decreases further, this peak
becomes a dominant feature. The peak amplitude reaches saturation
when the temperature decreases below 6-8 K, and at $T$ = 4.2 K, it
is almost 20 times as great as the conductivity observed at high
bias voltages. Note that this anomaly cannot be attributed to
Joule heating. The estimate of the sample heating for a typical
value of heat transfer to helium yields a conductivity decrease of
less than 10\% for the whole range of measured voltages. In
addition, the $I$-$V$ characteristics show a clearly defined set of
conductivity peaks that are symmetric about zero voltage and at
low temperature correspond to $|V|=$50, 100, and 150 mV, i.e., to
$|V|=nV_0$, where $V_0=$ 50 mV and $n$= 1,2, 3. As the temperature
increases above 25 K, these peaks move to lower energies, and, at
the temperature $T\backsimeq$59 K corresponding to the second
Peierls transition for NbSe$_3$, the values of $V_n$ vanish.

The picture described above was observed only for perfect
structures. The presence of structure defects, such as a twin
boundary, leads to a considerable reduction of the conductivity
peak at zero bias and to the appearance of a peak at $V=$ 25-27 mV
$\approx V_0/2$ (curve 2 in Fig. 3). Qualitatively similar
dependences were observed for some of the NbSe$_3$-NbSe$_3$ point
contacts. Their $I$-$V$ characteristics also exhibit a conductivity
peak at zero bias with an amplitude approximately equal to that
observed for defect overlap junctions, as well as a conductivity
peak at $V\approx$ 25 mV (curve 1 in Fig. 3). Figure 4 presents the
temperature dependences of the normalized dynamic conductivity at
zero bias voltage for all types of samples studied in the
experiment. One can see that, for a perfect overlap junction, the
amplitude of the zero bias anomaly is more than three times
greater than the amplitudes observed for the defect layered
structures and point contacts. Note that the voltage value $V_0=$
50 mV is close to twice the value of the low-temperature energy
gap, $2\Delta_{p2}/e$, for NbSe$_3$ \cite{10,11}, while the
dependence $V_n(T)/V_n(0)$ obtained for both overlap junctions and
point contacts (Fig. 5) agrees well with the temperature
dependence of the energy gap predicted by the BCS theory (the
dashed line in Fig. 5). This result suggests that the conductivity
features observed in the experiment are governed by a gap
mechanism.

It is important to note that all effects described above were
observed exclusively for the transport across the layers (along
the $a^*$ axis). The extra experiments performed by us on
specially fabricated bridges and point contacts oriented along the
$c$ axis (with the transport across the chains in the layer plane)
showed no conductivity peak at $V=0$.

Let us first analyze the results obtained for perfect overlap
junctions. The most prominent feature of the $I$-$V$ characteristics
of these structures is the strong conductivity peak at zero bias
voltage. In addition, the characteristics exhibit a periodic
sequence of peaks at $|V|=nV_0$, which resembles the series of
quasiparticle branches observed in the $I$-$V$ characteristics of
Bi-2212 layered structures when the measurements are performed
across the layers \cite{3}. As was noted above, the value $V_0=$
50 mV is close to twice the value of the low temperature CDW gap
in NbSe$_3$, and the temperature dependence $V_0(T)$ follows the
prediction of the BCS theory. However, in contrast to the case of
a superconductor, the conductivity observed for NbSe$_3$ at zero bias
voltage is finite. This fact suggests that the interlayer
conduction at zero bias does not originate from the collective CDW
contribution of the Josephson tunneling type.

The conductivity peak observed at zero bias cannot be explained by
the regular (incoherent) single-particle tunneling as well. If
this were the case, the conductivity peak could be attributed to
the energy dependence of the density of states of electrons that
are not condensed in the CDW. Then, the conductivity feature under
discussion should also be observed for N-I-CDW tunnel
junctions. However, no such effects were revealed by the detailed
experimental studies of this kind of tunnel junctions oriented
along the $a^*$ axis \cite{10,11}.

We believe that the conductivity peak at zero bias is caused by
the coherent interlayer tunneling of the charge carriers localized
in the pockets of the Fermi surface, and below we thoroughly
substantiate this statement.

In the general case, the tunnel current between two
nonsuperconducting layers, $a$ and $b$, is described by the expression
\cite{12}: \begin{equation} \label{one} I=\frac{4\pi
es}{\hbar}\sum_{\bf{p},{q}}|t_{ab}({\bf{p},{q}})|^2
\int_{-\infty}^{+\infty}d\omega
\left(\tanh\frac{\omega}{\textit{2T}}-\tanh\frac{\omega-\textit{eV}}{\textit{2T}}\right)
\mathrm{Im}G_a^R({\bf{p}},\omega)G_b^R({\bf{q}},\omega-\textit{eV}),
\end{equation}

where $G^R$ is the retarded Green function of an electron with the
momentum $\bf{p}$ in the layer $a$,
$\mathrm{Im}G^R(\bf{p},\omega)$ is the spectral density of the
Green function, $t_{ab}(\bf{p},\bf{q})$ is the matrix element
characterizing the tunneling from the state with the momentum
$\bf{p}$ in the layer $a$ to the state with the momentum $\bf{q}$
in the layer $b$, $V$ is the voltage between the adjacent layers,
and $T$ is the temperature.

Note that the contribution of the collective (moving) CDW mode
\cite{13} to the interlayer tunneling is absent, because the
tunnel current is determined by the Green functions in different
layers, whereas the collective mode propagates within the layers.

Below, we consider the coherent tunneling, when the electron
momentum in the plane does not change under the tunneling: $\bf{p}
= \bf{q}$. Then, we have

\begin{equation} \label{two}
I=\frac{es}{\pi\hbar}\int d{\bf{p}}|t_{ab}({\bf{p}})|^2
\int_{-\infty}^{+\infty}d\omega
\left(\tanh\frac{\omega}{\textit{2T}}-\tanh\frac{\omega-\textit{eV}}{\textit{2T}}\right)
\mathrm{Im}G_a^R({\bf{p}},\omega)G_b^R({\bf{p}},\omega-\textit{eV})
\end{equation}

 Taking into account that the
scattering in a layer is determined by the collision frequency
$\nu$, we obtain

\begin{equation} \label{three}
\mathrm{Im}G^R({\bf{p}},\omega)=\frac{\gamma}{\pi(\omega-\epsilon({\bf{p}})^2+\gamma^2)}
\end{equation}
where $\epsilon(\bf{p})$ is the electron spectrum and
$\gamma=\hbar\nu$. We can use this expression with
$\gamma=\gamma_{sc}+\gamma_{inc}$, which takes into account the
change in the momentum because of the scattering within the layers
and the change due to the tunneling
$\gamma_{inc}=\left<\epsilon({\bf{p}})-\epsilon({\bf{q}})\right>$. In both cases,
we take into account the energy uncertainty for the state
characterized by the momentum $\bf{p}$. We define the tunneling as
almost coherent when $\gamma_{inc}$ and $\gamma$ are small
relative to other energy parameters of the electron system. In our
case, such parameters are the Peierls gap and the width of the
electron band in the pockets. Replacing $t_{ab}(\bf{p})$ by the
momentum-independent quantity $t$ and integrating with respect to
$\bf{p}$, we obtain

\begin{equation} \label{four}
\frac{1}{4\pi^2}\int d{\bf{p}}|t_{ab}({\bf{p}})|^2
\mathrm{Im}G_a^R({\bf{p}},\omega)\mathrm{Im}G_b^R({\bf{p}},\omega-\textit{eV})
=\frac{2\gamma N(0)|t|^2}{e^2V^2+4\gamma^2}
\end{equation}
where $N(0)$ is the density of states of electrons in the pockets.
Finally, for the interlayer current when $eV<2\Delta$, we obtain
the expression

\begin{equation} \label{five}
I(V) =\frac{eV\gamma N(0)|t|^2}{2\pi^3(e^2V^2+4\gamma^2)}
\end{equation}
and the dynamic conductivity has the form

\begin{equation} \label{six}
\frac{\sigma(V)}{\sigma(0)}=4\gamma^2\frac{4\gamma^2-e^2V^2}{(e^2V^2+4\gamma^2)^2}
\end{equation}

Note that the temperature dependence is present in $\gamma(T)$
only. One can see that the dynamic conductivity has a peak of
width $\approx\gamma$ at $V$ = 0 and becomes negative and unstable
when $eV>2\gamma$. For a thirty-layer structure, the
experimentally observed peak width $\sim$10 mV corresponds to
$\gamma\approx$ 0.3 meV. For comparison, we estimate the parameter
$\gamma_{sc}$ characterizing the scattering within the layers. Using
the known mobility data $\mu=e/\nu_{sc}m^*\approx 4\times10^4$
cm$^2$/Vs \cite{7,14}, where $m^*=0.24m_e$ \cite{15}, we obtain
$\gamma_{sc}\approx0.13$ meV; i.e., the changes in the electron
momentum because of tunneling are approximately equal to changes
caused by the scattering within the layers.

Now, we explain why the interlayer current decreases with
increasing voltage V. The electron tunneling between the layers
must obey the energy conservation law; i.e., $\epsilon({\bf{p}})=
\epsilon({\bf{q}})-eV$ to within $2\gamma$. For coherent tunneling,
we have $\epsilon({\bf{p}})= \epsilon({\bf{q}})$, and for $eV\ll\gamma$,
tunneling is possible and we have the conventional Ohm's law. When
$eV>2\gamma$, tunneling is impossible up to the voltage $V$
reaching $2\Delta/e$. In this case, electrons condensed in the CDW
begin to contribute to the interlayer current in the form of a
regular tunneling of CDW quasiparticles through the double Peierls
gap $2\Delta$. Thus, the interlayer current can be realized by
means of only one of the two aforementioned mechanisms. In an
actual multilayer structure, because of the geometric
nonuniformity of individual layers (the areas of the layers are
somewhat different), the voltage value $V\sim 2\gamma/e$ will not
be reached simultaneously for different individual tunneling
junctions. When $V>2\gamma/e$, some of the junctions can be in the
coherent tunneling regime whereas other junctions can be in the
regime of single-particle tunneling through the gap. As a result,
a sequence of conductivity peaks must appear in the $I$-$V$
characteristic of the compound under study at the voltages
$V=2n\Delta_p/e$, where $n$ = 1, 2, $\dots\ $.

In the presence of defects in an overlap junction or in the case
of a point contact, the incoherence of tunneling is enhanced,
although in the best structures the tunneling remains almost
coherent. Therefore, the $I$-$V$ characteristics of these structures,
along with the conductivity peak at zero bias, contain an
additional feature at $V=\Delta_p/e$, which is related to the
single-particle tunneling of the N-I-CDW type.

Thus, the results of this study show that the interlayer tunneling
in natural layered structures obtained from NbSe$_3$ represents an
independent efficient method for investigating the Peierls state
in this compound. The tunneling conductivity peak observed at zero
voltage can be self-consistently explained by the almost coherent
interlayer tunneling of charge carriers that are not condensed in
the CDW and are localized in the Fermi surface pockets not covered
by the gap. The series of equidistant peaks in the $I$-$V$
characteristic can be explained by the quasiparticle tunneling
through the CDW gap under sequential transitions of individual
tunneling junctions to the resistive state. Note that the
conductivity peak observed at zero bias is a unique manifestation
of coherent single-particle transport in solids.

We are grateful to S.A. Brazovskii, V.A. Volkov, and V.M.
Yakovenko for useful discussions. This work was supported by the
Russian Foundation for Basic Research (project nos. 99-02-17364,
01-02-16321, and 00-02-22000 CNRS), the state program "Physics of
Solid Nanostructures" (project no. 97-1052), the NWO
Russian-Netherlands Project, and the Los Alamos National
Laboratory with the support of US DOE.

\newpage

\begin{figure}[t]
\includegraphics[width=7cm]{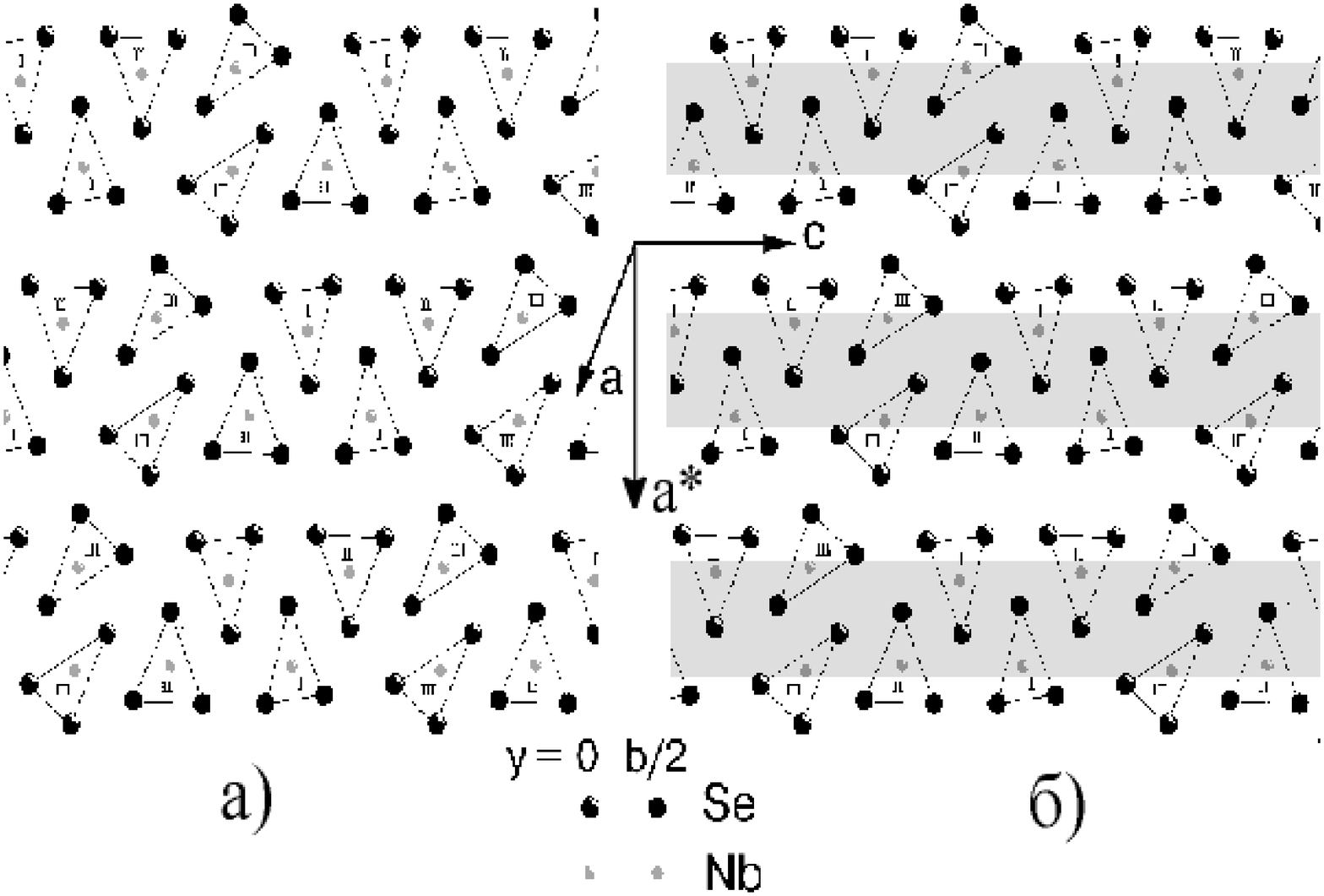}
\caption{ Schematic representation of the NbSe$_3$ structure in
the ($a-c$) plane; (b) the same structure with the conducting
planes indicated by shading} \label{Fi1}
\end{figure}

\begin{figure}[t]
\includegraphics[width=7cm]{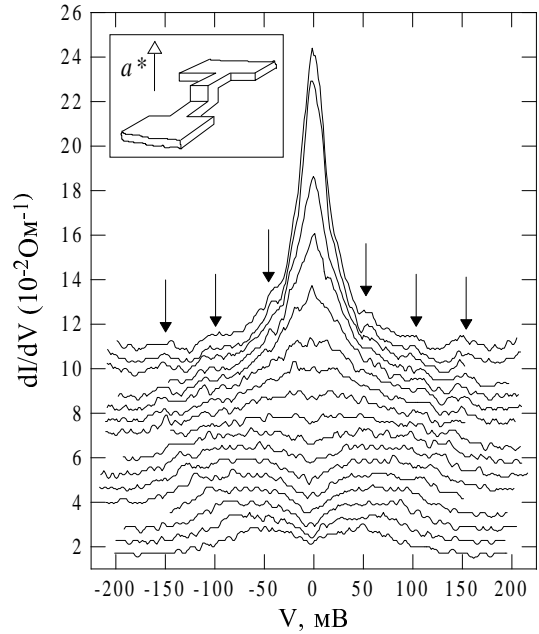}
\caption{Dependences of $dI/dV$ on the voltage $V$ for an overlap
junction (sample No. 3) at different temperatures: 59.5, 55.6,
52.1, 48.0, 43.7, 40.1, 37.0, 34.3, 31.0, 28.0, 25.1, 22.6, 19.1,
15.7, 12.8, 8.0, and 4.2 K. The dynamic conductivity scale
corresponds to the curve at $T$ = 59.5 K, and other curves are
shifted upwards for clarity. The inset shows the sample
configuration.} \label{Fi2}
\end{figure}

\begin{figure}[t]
\includegraphics[width=7cm]{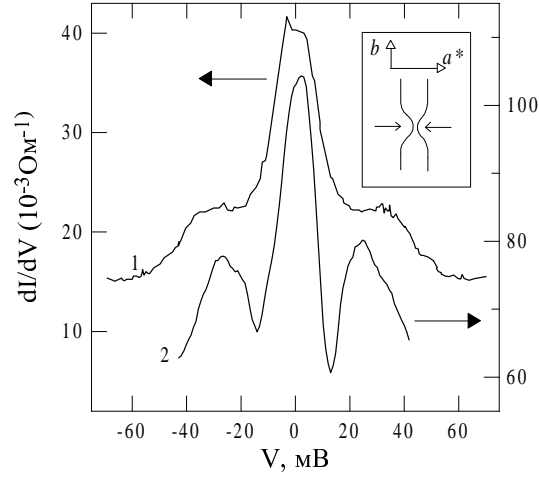}
\caption{Dependences $dI/dV(V)$ (1) for a NbSe$_3$-NbSe$_3$ point
contact and (2) for an overlap junction (sample No. 2) at $T$ =
4.2 K. The inset shows the point contact configuration.}
\label{Fi3}
\end{figure}

\begin{figure}[t]
\includegraphics[width=7cm]{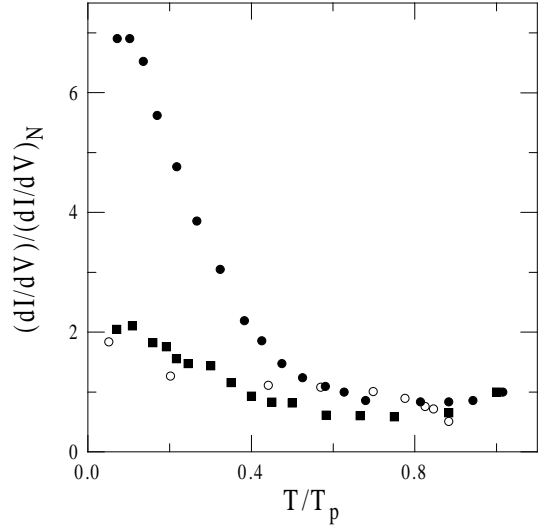}
\caption{Temperature dependences of the dynamic conductivity at
zero bias voltage for two overlap junctions: sample No. 3 (full
circles) and sample No. 1 (full squares), and for a
NbSe$_3$-NbSe$_3$ point contact (empty circles). The conductivity
values are normalized to the value of $dI/dV$ ($T$ = 62 K).}
\label{Fi4}
\end{figure}

\begin{figure}[t]
\includegraphics[width=7cm]{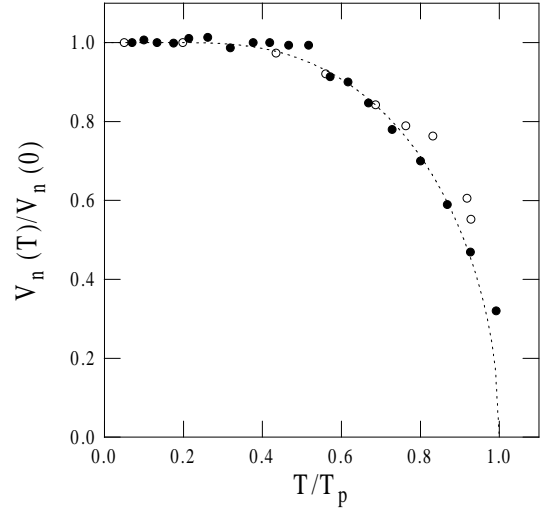}
\caption{Temperature dependence of the second Peierls energy gap
for an overlap junction (sample No. 3, full circles) and a
NbSe$_3$-NbSe$_3$ point contact (empty circles). } \label{Fi5}
\end{figure}

\end{document}